\documentclass[prd,aps,showpacs,preprintnumbers,amssymb]{revtex4}
\usepackage{graphicx}

\begin{document}

\title{%
\hfill{\normalsize\vbox{%
\hbox{\rm CP$^3$-Origins: 2009-12, SU-4252-898}
 }}\\
{ Perturbed S$_3$ neutrinos}}

\author{Renata Jora
$^{\it \bf a}$~\footnote[1]{Email:
 rjora@ifae.es}}

\author{Joseph Schechter
 $^{\it \bf b,c}$~\footnote[2]{Email:
 schechte@phy.syr.edu}}

\author{M. Naeem Shahid
$^{\it \bf b,c}$~\footnote[3]{Email:
   mnshahid@phy.syr.edu            }}

\affiliation{$^{\bf a}$ Grup de Fisica Teorica,
Universitat Autonoma de Barcelona, E -08193
Belaterra (Barcelona), Spain}

\affiliation{$^ {\bf \it b}$$CP{^3}$-Origins,
Campusvej 55, DK-5230 Odense M, Denmark}

\affiliation{$^ {\bf \it c}$ Department of Physics,
 Syracuse University, Syracuse, NY 13244-1130, USA,}

\date{\today}

\begin{abstract}
We study the effects of the perturbation which violates the
permutation symmetry of three Majorana neutrinos but
preserves the well known (23) interchange symmetry.
This is done in the presence of an arbitrary Majorana
phase $\psi$ which serves to insure the degeneracy of
the three neutrinos at the unperturbed level.
\end{abstract}


\maketitle

\section{Introduction}

    In the present paper, a particular approach  to understanding
    lepton mixing, proposed in \cite{jns} and further studied in \cite{cw},
    will be examined in more detail.
First, we briefly review the approach.

Of course,  the standard model interaction term for $\beta$ decay or $\pi^-
\rightarrow e^- \overline{\nu}_{e}$ includes the leptonic
piece:

\begin{equation}
   {\cal L}=
\frac{ig}{\sqrt{2}}W_{\mu}^-\overline{e}_{L}\gamma_{\mu}\nu_e
+h.c,
\label{Lbetadecay}
\end{equation}

The object $\nu_{e}$ is now known
\cite{superK}-\cite{minos} to be
 a linear combination
 of neutrino mass eigenstates, ${\hat \rho}_i$:
\begin{equation}
    \nu_{e}=\sum K_{ei}\widehat{\rho}_{i}
\label{mixing}
\end{equation}

where, in a basis with the charged leptons diagonal,
the full lepton mixing matrix is written as:
\begin{equation}
    K=\left(%
\begin{array}{ccc}
  K_{e1} & K_{e2} & K_{e3} \\
  K_{\mu1} & K_{\mu2} & K_{\mu3} \\
  K_{\tau1} & K_{\tau2} & K_{\tau3} \\
\end{array}%
\right)
\label{K}
\end{equation}
As has been discussed by many authors
 \cite{fx}-\cite{hvvm} the
 results of neutrino oscillation experiments
are (neglecting possible phases to be
discussed later)
 consistent with the
 ``tribimaximal mixing" matrix:
\begin{equation}
    K_{TBM}=\left(%
\begin{array}{ccc}
  \frac{-2}{\sqrt{6}} & \frac{1}{\sqrt{3}} & 0 \\
  \frac{1}{\sqrt{6}} & \frac{1}{\sqrt{3}} & \frac{1}{\sqrt{2}} \\
  \frac{1}{\sqrt{6}} & \frac{1}{\sqrt{3}} & \frac{-1}{\sqrt{2}} \\
\end{array}%
\right) \equiv R.
\label{R}
\end{equation}
Many different approaches have been used to explain
 the form of $K$.
A ``natural",and often investigated one
 uses the
parallel
three generation structure of the fundamental fermion
 families as a
starting point. An underlying discrete symmetry $S_{3}$,
the permutation group on three objects,
is then assumed. \cite{w}-\cite{mr}
The permutation matrices $S$ are,
\begin{eqnarray}
S^{(1)}&=& \left[
\begin{array}{ccc}
1&0& 0 \\
0&1&0\\
0&0&1
\end{array}
\right],\hspace{.3in} S^{(12)}= \left[
\begin{array}{ccc}
0&1& 0 \\
1&0&0\\
0&0&1
\end{array}
\right],\hspace{.3in}
 S^{(13)}= \left[
\begin{array}{ccc}
0&0&1 \\
0&1&0\\
1&0&0
\end{array}
\right],
\nonumber \\
S^{(23)}&=& \left[
\begin{array}{ccc}
1&0&0 \\
0&0&1\\
0&1&0
\end{array}
\right],\hspace{.3in}
 S^{(123)}= \left[
\begin{array}{ccc}
0&0&1 \\
1&0&0\\
0&1&0
\end{array}
\right],\hspace{.3in} S^{(132)}= \left[
\begin{array}{ccc}
0&1&0 \\
0&0&1 \\
1&0&0
\end{array}
\right],
\label{s3matrices}
\end{eqnarray}
This defining representation is not irreducible.
 The 3-dimensional space breaks
up into
 irreducible 2-dimensonal and 1-dimensional spaces.
One may note that the tribimaximal matrix, $K_{TBM}$
 is an example of the transformation
 which relates the
given basis to the irreducible one.
This fact provides our motivation
for further investigating the $S_3$ symmetry,
even though many other interesting approaches exist.
Of course, the symmetry requirement reads,
\begin{equation}
[S,M_{\nu}]=0, \label{commutator}
\end{equation}
where $S$ stands for any of the six matrices
 in Eq.(\ref{s3matrices}) and $M_{\nu}$
is the neutrino mass matrix.

 By explicitly evaluating the commutators one
obtains the solution:
\begin{equation}
M_\nu=\alpha
\left[
\begin{array}{ccc}
1&0&0\\
0&1&0\\
0&0&1
\end{array}
\right]+\beta
\left[
\begin{array}{ccc}
1&1&1\\
1&1&1\\
1&1&1\\
\end{array}
\right] \equiv \alpha {\bf 1}+\beta d .
\label{solution}
\end{equation}
$\alpha$ and $\beta$ are, in general, complex numbers
for the case of Majorana neutrinos
 while $d$ is
usually called the ``democratic" matrix.

 It is easy to verify
that this $M_\nu$ may be brought to diagonal
 (but not necessarily real) form
by
the real orthogonal matrix, $R=K_{TBM}$ defined above:
\begin{equation}
R^T(\alpha{\bf 1}+{\beta}d)R=\left[
\begin{array}{ccc}
\alpha&0&0 \\
0&\alpha+3\beta&0 \\
0&0&\alpha \\
\end{array}
\right].
\label{complexeigenvalues}
\end{equation}
 $R$ may be written in terms of the eigenvectors
 of $M_\nu$ as:
\begin{equation}
R= \left[
\begin{array}{ccc}
\vec{r}_1&\vec{r}_2&\vec{r}_3
\end{array}
\right],
\label{vecr}
\end{equation}
For example, $\vec{r}_1$ is the first column of
the tribimaximal matrix, Eq.(\ref{R}).
Physically one can assign different masses to the
 mass eigenstate
$\vec{r}_2$ in the 1-dimensional basis
 and to the (doubly
degenerate) eigenstates $\vec{r}_1$ and
$\vec{r}_3$ in the 2-dimensional basis.
At first glance this sounds ideal since it
 is well known that the
three neutrino masses are grouped into two
 almost degenerate
ones (``solar neutrinos") and one singlet,
with  different
values.
However, since we are demanding that R be taken as the
tribimaximal form, the physical identification requires
$\vec{r}_1$ and $\vec{r}_2$
 to be the
"solar" neutrino eigenstates rather than the
 degenerate ones
$\vec{r}_1$ and $\vec{r}_3$.
This had been considered a serious objection to the present
approach since often a scenario is pictured in which the mass
eigenvalue for $\vec{r_3}$ is considerably larger than
the roughly degenerate masses associated with
$\vec{r_1}$ and $\vec{r_2}$.
 A way out was
suggested in \cite{jns} where it was noted that,
 for
values of $m_1 + m_2 +m_3$ larger than around 0.3 eV,
 the neutrino spectrum
would actually be
approximately degenerate.
This may be seen in detail by consulting the chart
in Table 1 of \cite{jns} wherein the neutrino masses are
tabulated as a function of an assumed value of the third
neutrino mass, $m_3$. Actually it is seen that there is also a region
around $m_3\approx$ 0.04 eV
and $m_1+m_2+m_3 \approx 0.18 eV$
 where an assumed initial
degeneracy may be reasonable.
To make physical sense out of such a scenario,
 it was suggested that
 the neutrino mass
matrix be written as,
\begin{equation}
M_{\nu}=M_{\nu}^{(0)}+M_{\nu}^{(1)}+M_{\nu}^{(2)},
\label{pertM}
\end{equation}
where $M_{\nu}^{(0)}$ has the full $S_3$ invariance
and has degenerate (at least approximately) eigenvalues.
Furthermore, the smaller
 $M_{\nu}^{(1)}$ is invariant under a particular $S_2$
subgroup of $S_3$ and breaks the degeneracy.
Finally,
 $M_{\nu}^{(2)}$ is invariant
under a different $S_2$ subgroup of $S_3$ and is assumed
to be smaller still. The strengths are summarized as:
\begin{equation}
M_{\nu}^{(0)}>M_{\nu}^{(1)}>M_{\nu}^{(2)}.
\label{pertorders}
\end{equation}
This is inspired by the pre-QCD flavor perturbation theory
of the strong interaction which works quite well. In that
case the initially unknown strong interaction Hamiltonian
is expanded as
\begin{equation}
H=H^{(0)}+H^{(1)}+H^{(2)}.
\label{Hpert}
\end{equation}
Here $H^{(0)}$ is the dominant $SU(3)$ flavor invariant piece,
 $H^{(1)}$ is the smaller Gell-Mann Okubo perturbation
 \cite{gmo}
  which transforms
as the eighth component of a flavor octet representation and breaks
the symmetry to SU(2) and  $H^{(2)}$, which transforms as a different
component of the octet representation and breaks the symmetry further
to  the hypercharge U(1), is smaller still.

    There is a possible immediate objection to the assumption
that the neutrino mass eigenvalues be degenerate in the initial S$_3$
invariant approximation; after all Eq.(\ref{complexeigenvalues})
 shows that there are two
different eigenvalues $\alpha$ and $\alpha + 3 \beta$. This was
overcome by recognizing that these are both complex numbers and that
they could both have the same magnitude but different directions. Having
the same magnitude guarantees that all three physical masses will be the same.
This introduces a physical phase $\psi$ corresponding to the angle between
$\alpha$ and $\alpha + 3 \beta$.

    In the strong interaction case, the initial SU(3) invariance
was found to be reasonably well obeyed. It is thus natural to ask what
predictions may exist in the initial $S_3$ invariant approximation
in our neutrino model. It was found \cite{jns} that the leptonic factor for
neutrinoless double beta decay, $m_{ee}$ could be predicted in this
limit to be,
\begin{equation}
|m_{ee}|=\frac{m}{3}\sqrt{5+4cos{\psi}},
\label{ourmee}
\end{equation}
where $m$ is the degenerate neutrino mass and $\psi$\
is the Majorana type phase mentioned above.
This led to the inequality
\begin{equation}
m>|m_{ee}|\geq m/3.
\label{inequality}
\end{equation}

    The next step in the program is to consider the effect
of the perturbation  $M_{\nu}^{(1)}$. Many authors
\cite{mutau}-\cite{Gutmutau}
have suggested
that a $\mu$-$\tau$ symmetry ((23) symmetry in the present language)
is  associated with tribimaximal mixing in the neutrino sector.
Thus it is a natural $S_2$ symmetry choice for $M_{\nu}^{(1)}$.
Recently, Chen and Wolfenstein \cite{cw}
applied this type of perturbation to our present model with the
additional assumption that the Majorana phase $\psi$
takes the value $\pi$. This corresponds to CP conservation. Their
result for $|m_{ee}|$ is in agreement with the lower limit in
 Eq.(\ref{inequality}). Here
we will investigate the first perturbed case without assuming
that special value of $\psi$.

     Before going on to this we will present an amusing argument
to show that the (23) perturbation is naturally associated with the tribimaximal form
(modulo the majorana type phase $\psi$) rather than a tribimaximal form
multiplied by a rotation in the two dimensional degenerate subspace (which
is physically irrelevant at the $S_3$ invariant level).
This is based on the fact that
degenerate perturbation theory must be employed, which leads to a
stability condition.
Further we will show that
 other $S_2$ perturbations are mathematically
consistent but do not lead to the desired tribimaximal form.

\section{Effects of different perturbations}

   In the present framework there are three different
 possible perturbations, each
 characterized by the $S_2$ subgroup which
remains invariant.
   Let us first consider the favored perturbation which
 leaves invariant the
S$_2$ subgroup, consisting of $S^{(1)}$ and $S^{(23)}$.
Apart from a piece which may be reabsorbed in
 Eq.(\ref{solution}),
such a perturbation has the form,
\begin{equation}
\Delta=\left(%
\begin{array}{ccc}
  0 & 0 & 0 \\
  0 & t & u \\
  0 & u & t \\
\end{array}%
\right)
\label{23pert}
\end{equation}
where $t$ and $u$ are parameters.
It is convenient to adopt the language of
ordinary quantum mechanics perturbation
theory. We should then work in a basis like
Eq.(\ref{complexeigenvalues}) where
$M_\nu$ in Eq.(\ref{solution}) is diagonal.
However, because of the double degeneracy between
the eigenvectors $\vec{r}_1$ and  $\vec{r}_3$
in Eq.(\ref{R}), the matrix $R$
is not the unique one which diagonalizes $M_\nu$.
 We should really use the
more general matrix $RX(\xi)$ where $X(\xi)$
is given by:
\begin{equation}
X(\xi)=
\left(%
\begin{array}{ccc}
  cos\xi & 0 & -sin\xi \\
  0 & 1 & 0 \\
  sin\xi & 0 & \cos\xi \\
\end{array}%
\right).
\label{X}
\end{equation}
In this basis $\Delta$ has the form:
\begin{equation}
X^T R^T \Delta R X=
\left(%
\begin{array}{ccc}
  \frac{c^2(t+u)}{3}+s^2(t-u) & \frac{\sqrt{2}}{3}c(t+u) &
\frac{2sc}{3}(t-2u)\\
  \frac{\sqrt{2}}{3}c(t+u) & \frac{2}{3}(t+u) &
-\frac{\sqrt{2}}{3}s(t+u) \\
  \frac{2sc}{3}(t-2u) & -\frac{\sqrt{2}}{3}s(t+u) &
\frac{s^2(t+u)}{3}+c^2(t-u) \\
\end{array}%
\right).
\label{transformeddelta}
\end{equation}
Here, $c=\cos\xi$ and $s=\sin\xi$.
Note that, before adding a perturbation, the $S_3$
symmetry predicts the lepton mixing matrix to be
$RX(\xi)$ rather than the desired tribimaximal form,
$R$.

    In perturbation theory, the first correction
to the $m^{th}$ eigenvector involves the ratio
$\frac{<n|H^{(1)}|m>}{E_m-E_n}$. For degenerate
 perturbation theory
it is of course necessary that the numerator vanishes for
those states with $E_n=E_m$. Here we simply require
for the (13) matrix element:
\begin{equation}
(X(\xi)^T K_{TBM}^T \Delta K_{TBM} X(\xi))_{13}=0.
\label{stability}
\end{equation}
 This yields in general, $sin(2\xi)=0$. The
solution with $\xi=0$ is the desired tribimaximal form.
The solution with $\xi=\pi$ just changes the signs of the
first and third columns. However, the solutions with
$\xi=\pi/2$ and $\xi=3\pi/2$ interchange the first
 and third columns, which does not agree with experiment.
Thus, apart from a discrete ambiguity, the tribimaximal form
is uniquely chosen when a smooth connection with the (23)-type
 perturbation is required. Of course, the smooth connection
corresponds to choosing the correct initial states for
the perturbation treatment.

   It is easy to see that perturbations which leave
the other two $S_2$ subgroups invariant,
do not lead to  mixing matrices of
the desired tribimaximal form.
 The perturbation which commutes with $S^{(12)}$ is:
\begin{equation}
\Delta^{\prime}=\left(%
\begin{array}{ccc}
  t^{\prime} & u^{\prime} & 0 \\
  u^{\prime} & t^{\prime} & 0 \\
  0 & 0 & 0 \\
\end{array}%
\right).
\label{12pert}
\end{equation}
Similarly, the perturbation which commutes with
$S^{(13)}$ has the form:
\begin{equation}
\Delta^{\prime\prime}=\left(%
\begin{array}{ccc}
  t^{\prime\prime} &0& u^{\prime\prime}  \\
  0 & 0 & 0 \\
   u^{\prime\prime} & 0 & t^{\prime\prime} \\
\end{array}%
\right).
\label{13pert}
\end{equation}

The stability condition for
obtaining the tribimaximal mixing for
 the $\Delta^{\prime}$
pertubation would require the matrix element
$(K_{TBM}^T\Delta^{\prime}K_{TBM})_{13}$ to vanish;
 instead it
takes the value
$\frac{\sqrt{3}}{6}(t^{\prime}-2u^{\prime})$.
Similarly,
the stability condition for the $\Delta^{\prime\prime}$
pertubation does not work since the matrix element
$(K_{TBM}^T\Delta^{\prime\prime}K_{TBM})_{13}$
 takes the generally non-zero value
 $\frac{\sqrt{3}}{6}(-t^{\prime\prime}
+2u^{\prime\prime})$.

    While we have seen that the
 stability condition
for (23) invariant perturbations enforces
 the experimentally plausible tribimaximal mixing,
the underlying $S_3$ symmetry should allow
 characteristic stable mixing matrices
 to emerge for either the (12) invariant
or (13) invariant perturbations.
 What are their forms?
In the case of a (12) perturbation, the stability
 condition associated
with degenerate perturbation theory reads:
\begin{equation}
(K^T\Delta^{\prime}K)_{13}=0.
\label{12stability}
\end{equation}
Here the characteristic mixing matrix
 emerges as $K=K_{TBM}X(\xi)$ for a suitable
value of $\xi$.
The solution is easily seen to have the form:
\begin{equation}
K_{TBM}X(\frac{\pi}{6})=
\left(%
\begin{array}{ccc}
  \frac{1}{\sqrt{3}} & \frac{1}{\sqrt{2}}
 & \frac{1}{\sqrt{6}} \\
  \frac{1}{\sqrt{3}} & \frac{-1}{\sqrt{2}}
 & \frac{1}{\sqrt{6}} \\
  \frac{1}{\sqrt{3}} & 0 & \frac{-2}{\sqrt{6}} \\
\end{array}%
\right).
\label{X30}
\end{equation}
In the case of a (13) invariant perturbation,
 the stability
 condition associated
with degenerate perturbation theory reads:
\begin{equation}
(K^T\Delta^{\prime\prime}K)_{13}=0.
\label{13stability}
\end{equation}
Here the characteristic stable mixing matrix
 turns out to be:
\begin{equation}
K_{TBM}X(\frac{\pi}{3})=
\left(%
\begin{array}{ccc}
  \frac{-1}{\sqrt{6}} & \frac{1}{\sqrt{3}}
 & \frac{1}{\sqrt{2}} \\
  \frac{2}{\sqrt{6}} & \frac{1}{\sqrt{3}}
 & 0 \\
  \frac{-1}{\sqrt{6}} & \frac{1}{\sqrt{3}} &
\frac{-1}{\sqrt{2}} \\
\end{array}%
\right).
\label{X60}
\end{equation}

The situation is summarized in Table \ref{firsttable}.
Mathematically, any of the three perturbations
 will result
in a stable mixing matrix. However,
only the (23) perturbation gives the
experimentally allowed tribimaximal form.
For example, we see that the zero value of
$K_{13}$, in good present
agreement with experiment,
 only holds for the $\Delta$
[(23)-type] perturbation.

\begin{table}[htbp]
\begin{center}
\begin{tabular}{c|c}
\hline
  Perturbation & Mixing matrix  \\
\hline \hline
$\Delta$ & $K_{TBM}$       \\
$\Delta^{\prime}$ &  $K_{TBM}X(\frac{\pi}{6})$        \\
$\Delta^{\prime\prime}$ &  $K_{TBM}X(\frac{\pi}{3})$       \\
\hline
\end{tabular}
\end{center}
\caption[]{Characteristic, stable mixing matrices
for each $S_2$ invariant perturbation.
}
 \label{firsttable}
\end{table}

\section{Zeroth order setup}

\begin{figure}[htbp]
\centering
\rotatebox{0}
{\includegraphics[width=10cm,height=10cm,clip=true]{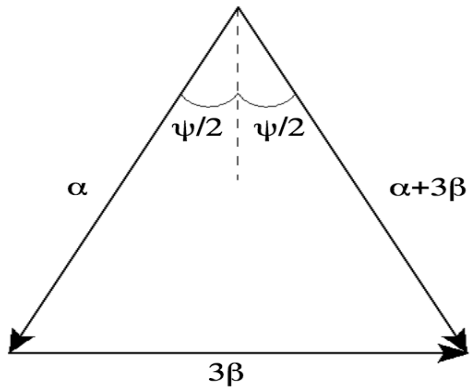}}
\caption[]{Isosceles triangle with angle $\psi$ between
the equal length 2-vectors $\alpha$ and $\alpha + 3\beta$.}
\label{triangle}
\end{figure}

    In order to go further we adopt convenient conventions
for the, in general, complex parameters $\alpha$ and $\beta$
defined in Eq.(\ref{solution}). The goal is to adjust a phase,
$\psi$ in order that the zeroth order spectrum has three exactly
degenerate neutrinos. As shown in Fig. \ref{triangle}, we take
the 2-vector 3$\beta$ to be real positive. Then the
 2-vector $\alpha$ lies in the third quadrant as:
\begin{equation}
\alpha=-i|\alpha|e^{-i\psi/2},
\label{alpha}
\end{equation}
where the physical phase $\psi$ lies in the range:
\begin{equation}
o<\psi\leq \pi.
\label{psirange}
\end{equation}
Finally $|\alpha|$ is related to $\beta$ by,
\begin{equation}
|\alpha|=\frac{3\beta}{2sin(\psi/2)}.
\label{alphabeta}
\end{equation}
In the limiting case $\psi=\pi$, $\alpha$ takes the real value,
\begin{equation}
\alpha=-\frac{3\beta}{2} \hspace{2cm}  (\psi=\pi).
\label{realalpha}
\end{equation}

\section{Analysis of favored perturbation}
For simplicity we will consider the parameters
$t$ and $u$ in Eq.(\ref{23pert}) to be real rather
than complex. The entire neutrino mass matrix to
first order is $M_\nu=\alpha{\bf 1} +\beta d +\Delta$.
Since we are working in a basis where the zeroth order
piece is diagonalized by the tribimaximal matrix, $R$,
we must diagonalize the matrix:
\begin{eqnarray}
&& R^T( \alpha{\bf 1} +\beta d +\Delta)R =
\nonumber \\
&& \alpha{\bf 1} +
\left(%
\begin{array}{ccc}
  t+u & \frac{\sqrt{2}}{3}(t+u) &
0\\
  \frac{\sqrt{2}}{3}(t+u) & 3\beta+ \frac{2}{3}(t+u) &
0 \\
  0 & 0 &
t-u \\
\end{array}%
\right).
\label{tobediag}
\end{eqnarray}

Diagonalizing the upper left 2 x 2 sub-matrix yields
the three, in general, complex eigenvalues:
\begin{eqnarray}
&& \alpha +\frac{3}{2}(\beta + T)
(1-\sqrt{1-\frac{4\beta T}{3(\beta+T)^2}})\approx \alpha +T,
\nonumber \\
&& \alpha +\frac{3}{2}(\beta + T)
(1+\sqrt{1-\frac{4\beta T}{3(\beta+T)^2}})\approx
\alpha +3\beta +2T,
\nonumber \\
&&\alpha +t -u,
\label{evs}
\end{eqnarray}
where we introduced the abbreviation,
$T=(t+u)/3$.
The indicated approximations to the exact
eigenvalues correspond to working to first
order in the parameters t and u. Remember
that according to our original setup, t and u
are supposed to be small compared to $|\alpha|$
and $\beta$. Since Fig. \ref{triangle} shows that
generally $|\alpha|> 3\beta/2$, it is sufficient that
$|t|$ and $|u|$ be small compared to $\beta$.

   In this approximation the corresponding eigenvectors
are the columns of,

\begin{equation}
R_1 \approx
\left(%
\begin{array}{ccc}
  1 & \frac{\sqrt{2}}{9\beta}(t+u) & 0 \\
  -\frac{\sqrt{2}}{9\beta}(t+u) & 1 &
0 \\
  0 & 0 &
1 \\
\end{array}%
\right).
\label{R1}
\end{equation}

The entire diagonalization may be presented as,
\begin{equation}
K^T(\alpha {\bf 1}+\beta d +\Delta)K =
\left(%
\begin{array}{ccc}
  m_1 & 0 & 0  \\
  0 & m_2 & 0 \\
  0 & 0 & m_3 \\
\end{array}%
\right).
\label{entirediag}
\end{equation}
Here $m_1$, $m_2$ and $m_3$ are the three (positive)
neutrino masses and
\begin{equation}
K=RR_1P
\label{fullmixingmat}
\end{equation}
is the full neutrino mixing matrix (in
a basis where the charged leptons are diagonal).
The neutrino masses, to order $(t,u)/\beta$, are
seen to be:
\begin{eqnarray}
m_1&\approx& \frac{3\beta}{2}csc(\frac{\psi}{2})\left[1-
\frac{2}{9\beta} (t+u)sin^2(\frac{\psi}{2})\right],
\nonumber \\
m_2&\approx& \frac{3\beta}{2}csc(\frac{\psi}{2})\left[1+
\frac{4}{9\beta} (t+u)sin^2(\frac{\psi}{2})
\right],
\nonumber \\
m_3&\approx& \frac{3\beta}{2}csc(\frac{\psi}{2})\left[1-
\frac{6}{9\beta} (t-u)sin^2(\frac{\psi}{2})\right].
\label{numasses}
\end{eqnarray}
These mass parameters were made real, positive
 by the introduction
of the phase matrix:
\begin{equation}
P=
\left(%
\begin{array}{ccc}
  e^{-i\tau} & 0 & 0  \\
  0 & e^{-i\sigma} & 0 \\
  0 & 0 & e^{-i\rho} \\
\end{array}%
\right),
\label{phasematrix}
\end{equation}
where,
\begin{eqnarray}
\tau&\approx&  \frac{\pi}{2}+\frac{1}{2}tan^{-1}[
\frac{cot(\psi/2)}{1-\frac{2(t+u)}{9\beta}}]
\nonumber \\
\sigma&\approx& \pi-\frac{1}{2}tan^{-1}[
\frac{cot(\psi/2)}{1+\frac{4(t+u)}{9\beta}}]
\nonumber \\
\rho&\approx&  \frac{\pi}{2}+\frac{1}{2}tan^{-1}[
\frac{cot(\psi/2)}{1-\frac{2(t-u)}{3\beta}}].
\label{majphases}
\end{eqnarray}

To compare with experiment, we have important
 information from neutrino oscillation
 experiments \cite{superK}-\cite{minos}.
  It is known that \cite{A}
\begin{eqnarray}
A&\equiv& m_2^2-m_1^2= (8\pm 0.3)\times 10^{-5} eV^2,
\nonumber \\
B&\equiv& |m_3^2-m_2^2|= (2.5\pm 0.5)\times 10^{-3} eV^2.
\label{AB}
\end{eqnarray}
Also, constraints on cosmological structure formation yield
\cite{cosmobound} a
rough bound,
\begin{equation}
m_1+m_2+m_3 < 0.7 eV .
\label{cosmo}
\end{equation}
The two allowed spectrum types are:
\begin{eqnarray}
&&Type 1:\hspace{1cm} m_3>m_2>m_1,
\nonumber \\
&&Type 2:\hspace{1cm} m_2>m_1>m_3.
\label{spectrumtype}
\end{eqnarray}.

Now, from Eq.(\ref{numasses}) we see to leading order:
\begin{eqnarray}
 m_2^2-m_1^2&=&3\beta(t+u),
\nonumber \\
 m_3^2-m_2^2&=&\beta(-5t+u).
\label{deltanus}
\end{eqnarray}

The quantities $\beta t$ and $\beta u$
may thus be obtained for a type 1 spectrum as:
\begin{eqnarray}
&&\beta t = A/18 -B/6 \approx -4.13 \times 10^{-4} eV^2,
\nonumber   \\
&&\beta u =5A/18 + B/6 \approx 4.39 \times 10^{-4} eV^2,
\label{solve1}
\end{eqnarray}
where the central experimental values were used.
In the type 2 spectrum case, we should change $B \rightarrow
-B$ in the above to find,
\begin{eqnarray}
&&\beta t = A/18 +B/6 \approx 4.21 \times 10^{-4} eV^2,
\nonumber   \\
&&\beta u =5A/18 - B/6 \approx -3.94 \times 10^{-4} eV^2
\label{solve2}
\end{eqnarray}

Thus the, assumed real,  $S_3$ violation parameters $\beta t$ and $\beta u$
are now known for each spectrum type. Information about the quantity $\beta$
may in principle be obtained from the perturbed lepton mixing
matrix given in Eq. (\ref{fullmixingmat}):
\begin{equation}
K\approx
\left(%
\begin{array}{ccc}
  \frac{-2}{\sqrt{6}}-\frac{\sqrt{2}(t+u)}{9\beta\sqrt{3}} &
\frac{1}{\sqrt{3}}-\frac{2(t+u)}{9\beta\sqrt{3}} & 0 \\
  \frac{1}{\sqrt{6}}-\frac{\sqrt{2}(t+u)}{9\beta\sqrt{3}}&
\frac{1}{\sqrt{3}} +\frac{(t+u)}{9\beta\sqrt{3}}
& \frac{1}{\sqrt{2}} \\
  \frac{1}{\sqrt{6}}-\frac{\sqrt{2}(t+u)}{9\beta\sqrt{3}} &
\frac{1}{\sqrt{3}} +\frac{(t+u)}{9\beta\sqrt{3}}
 & \frac{-1}{\sqrt{2}} \\
\end{array}%
\right)P .
\label{perturbedK}
\end{equation}

With a usual parameterization \cite{mns} the matrix with zero (13) element
takes the form,

\begin{equation}
K=
\left(%
\begin{array}{ccc}
  c_{12} &               s_{12}& 0 \\
  -s_{12}c_{23}&
c_{12}c_{23}
& s_{23} \\
  s_{12}s_{23}&
-c_{12}s_{23}
 & c_{23} \\
\end{array}%
\right)P ,
\label{usuparam}
\end{equation}
where $c_{12}$ is short for $cos\theta_{12}$ for example.
This amounts to the
 predictions,
\begin{eqnarray}
c_{12}= -\frac{2}{\sqrt{6}}-\frac{\sqrt{2}(\beta
t+\beta u)}{9\sqrt{3}{\beta}^2},
\nonumber  \\
c_{23}= -\frac{1}{\sqrt{2}},
\nonumber  \\
s_{13}=0.
\label{c12dev}
\end{eqnarray}
Notice that, when the perturbation is absent, this agrees with
the tribimaximal form used here if both $\theta_{12}$ and $\theta_{23}$
lie in the second quadrant.
The results of a recent study
\cite{stv}
of neutrino oscillation experiments
are:
\begin{eqnarray}
(s_{23})^2&=& 0.50^{+0.07}_{-0.06},
\nonumber   \\
(s_{12})^2 &=& 0.304^{+0.022}_{-0.016}.
\label{exptangles}
\end{eqnarray}
One immediately notices that the prediction,
 $(s_{23})^2$=1/2 is unchanged from its
 tribimaximal value by the perturbation and agrees with the new
 analysis. On the other hand the tribimaximal prediction,
 $(s_{12})^2$=1/3 is slightly changed from its tribimaximal value
 and actually lies slightly above the upper experimental error bar.
This is probably not a serious disagreement but  it might be
instructive to
 try to fix it
using the predicted perturbation in the present model:
\begin{equation}
(s_{12})^2 =\frac{1}{3} -\frac{4}{27}\frac{\beta t +\beta u}{\beta^2}.
\label{perturbs12}
\end{equation}
For either the type 1 or type 2 assumed spectrum, the
perturbation is seen to be in the correct direction to lower
the value of  $s_{12})^2$, as desired. However, because
of the large cancellation between $\beta t $ and $\beta u$, this
effect is extremely small for a reasonable value
of $\beta^2$; even with $\beta$ as small as 0.05 eV,
 $(s_{12})^2$ is only lowered to 0.332.

\begin{figure}[htbp]
\centering
\rotatebox{0}
{\includegraphics[width=10cm,height=10cm,clip=true]{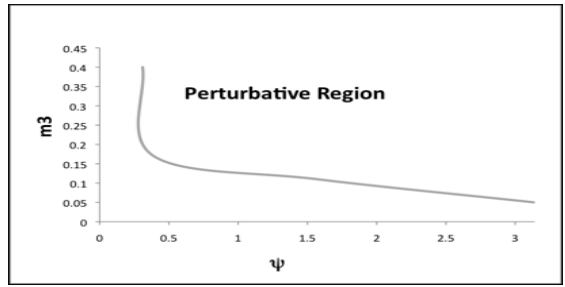}}
\caption{Sketch of perturbative region  in the $m_3$-$\psi$ plane. It is
about the same for both type 1 and type 2 neutrino spectra.
Note that $\psi$ is measured in radians and $m_3$ is measured in eV.}
\label{pregion}
\end{figure}

    It is also interesting to discuss the absolute masses
    of the neutrinos rather than just the differences of their squares.
    Since the differences are known, let us focus on
    one of them, say $m_3$:
     \begin{equation}
m_3 \approx \frac{3\beta}{2}csc(\frac{\psi}{2})
-\frac{\beta t -\beta u}{\beta}sin(\frac{\psi}{2}).
\label{m3}
\end{equation}
Notice that the first term on the right hand side is,
using Eq.(\ref{alphabeta}), simply the zeroth order
degenerate mass, $|\alpha|$ while the second term
represents the correction. Also note that (see Fig. \ref{triangle})
the point $\psi=0$ is not allowed. Considering $\psi$ as
a parameter (related to the strength of neutrinoless double beta decay),
this equation represents a quadratic formula giving $\beta$ in
terms of the absolute mass $m_3$ for any assumed $\psi$.
In Fig. \ref{pregion}, adopting the criterion that  $|t|/\beta$ and $|u|/\beta$ be
less than 1/5 for perturbative behavior, we display the perturbative region
in the $m_3-\psi$ plane.  In contrast to the case of $m_3$, $m_1$ and $m_2$
are seen to have small
 corrections since they of course depend on
$\beta(t+u)$ rather than $\beta(t-u)$.

\section{Neutrinoless double beta decay}

The characteristic physical novelty of of a theory with Majorana type
neutrinos is the prediction of a small, but non-zero, rate  for the
neutrinoless double beta decay of a nucleus: (A,Z) $\rightarrow$
(A,Z+2) + 2$e^-$. The appropriate leptonic factor describing the amplitude for
this process is,

\begin{equation}
|m_{ee}|= |m_1(K_{11})^2 + m_2(K_{12})^2 + m_3(K_{13})^2|.
\label{mee}
\end{equation}
 Substituting in the neutrino masses to order $(t,u)/\beta$ from
 Eq.(\ref{numasses} as well as Eq.(\ref{phasematrix}) yields:
\begin{equation}
|m_{ee}| \approx \frac{3\beta}{2sin(\frac{\psi}{2})}|\frac{2}{3}
+\frac{4(t+u)}{27\beta}cos^2(\frac{\psi}{2})
+[\frac{1}{3}-\frac{4(t+u)}{27\beta}cos^2(\frac{\psi}{2})]
e^{2i(\tau-\sigma})|.
\label{meetwo}
\end{equation}
The needed intermediate quantity $cos[2(\tau-\sigma)] $may
be easily obtained from Eqs.(\ref{majphases}) by construction of a suitable
right triangles to be:
\begin{equation}
cos[2(\tau-\sigma)] \approx cos \psi -\frac{(t+u)sin\psi}{9\beta}.
\nonumber
\end{equation}

We then find, correct to first order in $(t,u)/\beta$,
\begin{equation}
|m_{ee}| = \frac{\beta}{2sin(\psi /2)}\sqrt{5+4cos\psi},
\label{meethree}
\end{equation}
which is just the zeroth order result.
The experimental bound on $|m_{ee}|$ is given \cite{kka} as,

\begin{equation}
|m_{ee}| < (0.35 - 1.30) eV,
\label{meebound}
\end{equation}

which is small enough so that there is hope  the
possibility of a Majorana neutrino might be settled in the near future.
Since the correction to $|m_{ee}|$ has been seen to be zero
in this model we can take over the zeroth order inequality in
Eq.(\ref{inequality}). This means that the existence of the Majorana
phase, $\psi$ can alter the amplitude
for neutrinoless double beta decay by a factor of three
for given (approximately degenerate) neutrino masses.

\section{Summary and discussion}

    In some ways the problem of  ``flavor" in the Standard Model is reminiscent of
    that  in Strong Interaction  physics  before the quark model. At that time it was
    realized that, as a precursor to detailed dynamics, group theory might give important clues.

    Then the
    strong interactions were postulated to be SU(3) flavor invariant with a weaker
    piece having just the the SU(2) isospin (times hypercharge) invariance.
    In addition it was known that there was a still weaker isospin breaking (possibly QED)
    which by itself preserved a different SU(2) invariance (so-called U-spin).

    Here, an analogy for neutrinos of the first two steps was studied in a
    perturbative framework. In \cite{jns} and \cite{cw} the possibility that
    the neutrinos are not strictly degnerate at the unperturbed level
     was contemplated. However in this
    paper we have examined a strictly degenerate first stage (setting to zero the
     parameters called respectively $\epsilon$ and $b$ in those two papers).

     At the $S_3$ invariant level the neutrino mixing matrix is actually arbitrary
     up to a rotation in a 2-dimensional subspace. This problem can be settled
     (since degenerate perturbation theory is involved) by
     specifying the transformation property  of the perturbation to be added.
     Although there is widespread agreement that the first perturbation
     should preserve the $S_2$ subgroup which interchanges the second and third neutrinos,
     we presented in section II, for completeness and interest, the mixing matrices for
      the other two possibilities also.

      In sections III and IV we carried out the perturbation analysis for any choice of a
       Majorana-type phase, $\psi$ which plays an important role in this model. If $\psi$
       is considered fixed there are three parameters in the model (In \cite{cw}   $\psi$
       was considered fixed at the value $\pi$.) These three parameters can be taken as
       $\beta t$, $\beta u$ and $ \beta$ defined above. The quantities
       $\beta t$ and $\beta u$ were found in terms of the neutrino squared mass
       differences for each choice of neutrino spectrum type, ie normal or inverted hierarchy. The
       value of $\beta$ depends on the presently unknown absolute value of any neutrino mass.
       The magnitudes of $\beta t$ and $\beta u$ are similar (though not exactly equal) but differ
       in sign. Thus the perturbation corrections which involve ($\beta t $ + $\beta u$) are very
       small. Clearly (see the first of Eqs.(\ref{deltanus})) this is due to the small solar neutrino
       mass difference.
       This situation occurs for the correction to the mixing parameter $sin^2 \theta_{12}$ in
       addition to
       $m_1$ and $m_2$, the masses of the first two neutrinos. The perturbation dependence
       on ($\beta t $ -$\beta u)$ is not suppressed however. This occurs for the mass, $m_3$
       of the third neutrino. This result was used to make a sketch of the region in the
       $\psi$-$m_3$ plane for which the perturbation approach given seems numerically
       reasonable.

       The explicit role of the Higgs sector, which is believed to be at the heart
       of the matter, was not discussed in the present paper. However, this as well as some
        further technical details were discussed in \cite{jns}. Further treatment of this aspect is
        interesting for future work as is a detailed  investigation of the weakest perturbation, the analog
       of  the  U-spin preserving perturbation in the strong interaction. This could be used to further study
       other consequences \cite{cw} of possibly non-zero $\theta_{13}$.

\section{acknowledgments}

    We are happy to thank Amir Fariborz, Salah Nasri and Francesco Sannino for helpful
    discussions and encouragment.
    The work of J. Schechter and M.N. Shahid was supported in part by
    the US DOE under Contract No. DE-FG-02-85ER 40231; they would also like
    to thank
    the CP$^3$-Origins group at the University of Southern Denmark for their warm
    hospitality and partial support.


\begin{thebibliography}{9}

\bibitem{jns}R. Jora, S. Nasri and J. Schechter, Int. J. Mod. Phys. A,
{\bf 21}, 5875 (2006).

\bibitem{cw}C.-Y. Chen and L. Wolfenstein, Phys. Rev. D {\bf 77},
093009 (2008).

\bibitem{superK}Super Kamiokande collaboration, S. Fukuda et al,
Phys. Lett. B {\bf 539}, 179 (2002), hep-ex/0205075.

\bibitem{kamland}KamLAND collaboration, K. Eguchi et al, Phys. Rev.
Lett. {\bf 90}, 021802 (2003).

\bibitem{sno}SNO collaboration, Q. R. Ahmad et al,nucl-ex/
0309004.

\bibitem{k2k}K2K collaboration, M. H. Ahn et al, Phys. Rev. Lett. {\bf
90}, 041801 (2003).

\bibitem{gall}GALLEX Collaboration, W. Hampel et al, Phys. Lett. B
{\bf 447}, 127 (1999).

\bibitem{sage}SAGE Collaboration, J. N. Abdurashitov et al,
Phys. Rev. C {\bf 60}, 055801 (1999).

\bibitem{chooz}CHOOZ Collaboration, M. Apollonio et al,
Eur. Phys. J. C {\bf 27}, 331 (2003), hep-ex/0301017.

\bibitem{minos}MINOS Collaboration, Phys. Rev. D {\bf 73}, 072002
 (2005), hep-ex/0512036.

\bibitem{fx}H. Fritzsch and Z.-Z.Xing, Phys. Lett. B {\bf 440},
313 (1988), hep-ph/9808272.

\bibitem{HPS}P. F. Harrison, D. H. Perkins and W. G. Scott, Phys.
Lett. {\bf B530},79 (2002), hep-ph/0202074.

\bibitem{X}Z.-Z.Xing, Phys. Lett. B {\bf 533}, 85 (2002),
 hep-ph/020409.

\bibitem{HZ}X.G.He and A. Zee, Phys. Lett. B {\bf 560}, 87 (2003),
hep-ph/0204049.

\bibitem{HS} P. F. Harrison and W. G. Scott, hep-ph/0302025.

\bibitem{LV}C.I.Low and R.R.Volkas, Phys. Rev. D {\bf 68},
033007 (2003), hep-ph/0305243.

\bibitem{Z}A.Zee, Phys. Rev. D {\bf 68}, 093002 (2003),
hep-ph/0307323.

\bibitem{BHS} J.D. Bjorken, P. F. Harrison and W. G. Scott,
hep-ph/0511201.

\bibitem{MNY} R.~N.~Mohapatra, S.~Nasri and H.~B.~Yu,
  arXiv:hep-ph/0605020.

\bibitem{king} S.~F.~King, Nucl.\ Phys.\ B {\bf 576}, 85 (2000);
   S.~F.~King and N.~N.~Singh, Nucl.\ Phys.\ B {\bf 591}, 3 (2000).

\bibitem{mamain} E. Ma, Phys. Rev. D {\bf 70} 091301 (2004).

\bibitem{hvvm}M. Hirsch, A. Velanova del Morel, J.W.F. Valle and
E. Ma, Phys. Rev. D {\bf 72}, (031901) (2005).

\bibitem{w}L. Wolfenstein, Phys. Rev. D {\bf 18}, 958 (1978).

\bibitem{ps} S. Pakvasa and H. Sugawara,
Phys. Lett. B {\bf 73}, 61 (1978);
 {\bf 82}, 105 (1979); E. Derman and H.S.Tsao, Phys.
 Rev. D {\bf 20}, 1207 (1979) and Y. Yamanaka, H. Sugawara
and S. Pakvasa Phys. Rev. D {\bf 25}, 1895 (1982).

\bibitem{cfm}S.-L. Chen, M. Frigerio and E. Ma, hep-ph/0404084.

\bibitem{fty}M. Fukugita, M. Tanimoto and T. Yanagida,
Phys. Rev. D {\bf 57}, 4429 (1998), hep-ph/9709388.

\bibitem{mr}E. Ma and G. Rajasekaran, Phys. Rev. D {\bf 64},
113012 (2001), hep-ph/0106291.
----
\bibitem{gmo}M. Gell-Mann, Phys. Rev. {\bf 125}, 1067 (1962);
S. Okubo, Prog. Theor. Phys. {\bf 27}, 949 (1962); {\bf 28},
 24 (1962).

\bibitem{mutau} T. Fukuyama and H. Nishiura, hep-ph/9702253;
R. N. Mohapatra and S. Nussinov, Phys. Rev. {\bf D 60}, 013002
(1999); E. Ma and M. Raidal, Phys. Rev. Lett. {\bf 87}, 011802
(2001); C. S. Lam, hep-ph/0104116; T. Kitabayashi and M. Yasue,
Phys.Rev. {\bf D67} 015006 (2003); W. Grimus and L. Lavoura,
hep-ph/0305046; 0309050; Y. Koide, Phys.Rev. {\bf D69}, 093001
(2004);Y. H. Ahn, Sin Kyu Kang, C. S. Kim, Jake Lee, hep-ph/0602160;
 A. Ghosal, hep-ph/0304090; W. Grimus and L. Lavoura, Phys.\ Lett.\ B {\bf 572}, 189 (2003);
W.~Grimus and L.~Lavoura, J.\ Phys.\ G {\bf 30}, 73 (2004).


\bibitem{moh}  W. Grimus, A. S.Joshipura, S. Kaneko, L.
Lavoura, H. Sawanaka, M. Tanimoto, hep-ph/0408123; R. N. Mohapatra,
JHEP, {\bf 0410}, 027 (2004); A. de Gouvea, Phys.Rev. {\bf D69},
093007 (2004); R. N. Mohapatra and W. Rodejohann, Phys. Rev. {\bf D
72}, 053001 (2005); T. Kitabayashi and M. Yasue, Phys. Lett,. {\bf B
621}, 133 (2005);  R.~N.~Mohapatra and S.~Nasri, Phys.\ Rev.\ D {\bf
71}, 033001 (2005);R.~N.~Mohapatra, S.~Nasri and H.~B.~Yu, Phys.\
Lett.\ B {\bf 615}, 231 (2005).

\bibitem{Gutmutau}  K. Matsuda and H. Nishiura,
 Phys.\ Rev.\ D {\bf 73}, 013008 (2006);  A. Joshipura, hep-ph/0512252;
R. N. Mohapatra, S. Nasri and H.~B.~Yu, Phys. Lett. {\bf B 636}, 114
(2006).

\bibitem{A} C. Amsler et al, Review of Particle Physics, Phys. Lett.,
B667:1 (2008).

\bibitem{cosmobound}D. N. Spergel et al, Astrophys. J. Suppl.
{\bf 148}: 175 (2003);
S. Hannestad, JCAP {\bf 0305}: 004 (2003).

\bibitem{mns} See, for example, Eq. (10) of S.S. Masood, S. Nasri
and J. Schechter, Phys. Rev. D {\bf 71}, 093005 (2005). This
reference also discusses a more symmetrical parameterization
which may be convenient  for treating
neutrinoless double beta decay in general.


\bibitem{stv} T. Schwetz, M. Tortola and J.W.F. Valle,
arXiv:0808.2016.


\bibitem{kka}H. V. Klapdor-Kleingrothaus {\it et al}, Eur. Phys. J. A
{\bf 12}, 147 (2001). See also the review, C. Aalseth
{\it et al}, arXiv:hep-ph/0412300.


\end{thebibliography}
\end{document}